# On the spin and orbital variability of the intermediate polars

V. Breus[1], I.L. Andronov[1], P. Dubovsky[2], K. Petrik[3,4], S. Zola[5,6]

[1] Department of Mathematics, Physics and Astronomy, Odessa National Maritime University, Odessa, Ukraine
[2] Vihorlat Astronomical Observatory, Humenne, Slovak Republic
[3] Hlohovec Astronomical Observatory, Hlohovec, Slovak Republic
[4] Institute of Physics, Silesian University in Opava, Czech Republic
[5] Astronomical observatory of the Jagiellonian University, Krakow, Poland
[6] Mt. Suhora Observatory, Pedagogical University, Krakow, Poland

*Abstract*

We present a review of the results of long-term photometric monitoring of selected magnetic cataclysmic binary systems, which belong to a class named "Intermediate polars". We found a spin period variability in the V2306 Cygni system. We confirm the strong negative superhump variations in the intermediate polar RX J2133.7+5107 and improved a characteristic time of white dwarf spin-up in this system. We have investigated the periodic modulation of the spin phases with the orbital phase in MU Camelopardalis. We can propose simple explanation as the influence of orbital sidebands in the periodic signal produced by intermediate polar.

*Introduction*

Cataclysmic variables are close binary systems consisting of a white dwarf and a main-sequence star filling the Roche lobe. The gravity of the primary component leads to capture of matter from the secondary component near the inner Lagrangian point. Due to the high angular momentum of the plasma leaving this point, the stream can not be accreted directly by the compact star, and instead it forms an accretion disk around the white dwarf.

The subclass of magnetic cataclysmic binary systems has a primary component with strong magnetic field that can destroy the inner part of the accretion disk or prevent it from being formed. Depending on the strength of the magnetic field, these systems are divided into intermediate polars and polars.

The magnetic field of the primary component of the intermediate polars, often called DQ Her type stars, destroys the inner part of the accretion disk and matter is being accreted along the magnetic field lines, leading to the formation of one or two accretion columns near the magnetic poles. The matter forms a shock wave, heats up, and settles down on the surface of the white dwarf. The accretion columns in such systems are often the brightest sources of polarized radiation in a wide spectral range from X-ray to radio.

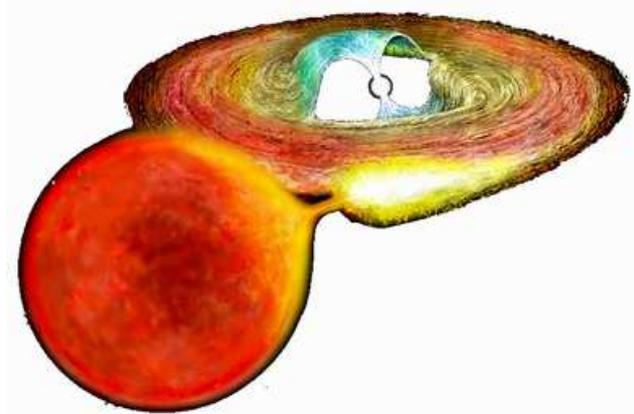

Fig. 1: Schematic picture of intermediate polar: white dwarf, red dwarf, accretion disk and 2 accretion columns.

Usually intermediate polars show two types of optical variability. The orbital variability is caused by the rotation of the system. During one revolution, we see the emission sources i.e. stars, an accretion disk and a hot spot on the disk at different angles. Typically, the orbital period of intermediate polar is about 3-7 hours.

The spin variability is caused by the rotation of the white dwarf with one or two accretion columns, with the periods ranging from few to dozens of minutes.

Flickering, rare outbursts, changes from high to low luminosity states and other processes are also observed in this kind of objects, therefore, the light curve is a superposition of periodic and aperiodic processes of different matter.

The most comprehensive review was presented by Warner [1]. Cataclysmic variables with different degree of influence of the magnetic field onto accretion were reviewed e.g. by Andronov [27].

*Observations and data reduction*

Long term monitoring of intermediate polars is performed as a part of the Inter-Longitude Astronomy campaign [2] and "Ukrainian Virtual Observatory" project [3].

High quality and long time series observations allow us to investigate fine effects on complex light curves and study the variations of the white dwarf rotation that cause period variations in intermediate polars.

We regularly obtain photometric data using Vihorlat National Telescope at the Astronomical Observatory on Kolonica Saddle, Slovakia (diameter of the main mirror is 1 meter); 60 cm Zeiss Cassegrain telescope at the Observatory and Planetarium of M. R. Stefanik in Hlohovec, Slovakia; 50 cm Zeiss and 40 cm Maksutow telescopes of Fort Skala Astronomical Observatory of the Jagiellonian University in Krakow, Poland. Sometimes we obtain time series from other telescopes in Ukraine, Korea, Hungary, Slovakia. In our research we use also long CCD time series from AAVSO international database and data obtained by such projects like ASAS, SuperWASP and others.

The reduction, consisting of calibration of scientific images for bias, dark and flatfield and extraction of instrumental magnitudes, was carried out with the Muniwin[4] and CoLiTecVS[5] software packages. The final derivation of magnitudes was obtained using the multiple comparison stars method described by

Kim et al. [6] and implemented in Multi-Column View[7] (MCV) by I.L. Andronov and A.V. Baklanov. Period analysis and determination of extrema timings were carried out using Variable Stars Calculator [8,9] (VSCalc) and MCV.

The (O-C) analysis was performed to study the variability of the orbital and spin periods of the systems. To increase the accuracy, for some data sets we determined one extremum per night of observations, for other we joined few consequent nights instead of using individual extrema timings. The method has been previously widely used for approximation of observations of intermediate polars (see Andronov & Breus [10]) and its last modifications including multiple iterations of (O-C) analysis for improvement of the value of the period of the system was recently described by Breus et al. [11].

As an (O-C) diagram we used to analyze the dependence of phase on the Julian date or the integer cycle number. The linear trend on this diagram argues for the necessary period correction, the parabolic trend shows a presence of period changes (in case of spin period of white dwarf in intermediate polars – spin-up or spin-down of this compact star). More complicated changes are frequently observable: spin-down may change to spin-up and back, sometimes the best fit for the O-C is a superposition of parabolic and periodic trigonometric functions.

We also study asynchronous polars, in which the rotation of the white dwarf is nearly synchronous with the orbital motion of the system [28], [29], [30] and classical polars, e.g. AM Her [31].

*V2306 Cygni*

The pulsating X-ray source 1WGAJ1958.2+3232 was discovered by Israel et al. in 1998 [12]. The star was named V2306 Cyg in 2003. Zharikov et al. [13] reported the detection of the orbital period of $4^h36^m$ ($0^d.1802\pm0^d.0065$) from photometry and the final value of $0^d.18152\pm0^d.00011$ from radial velocity variations. They confirmed the presence of (733.82±1.25) seconds spin period of the white dwarf using the spectroscopy and photometry and interpreted strong modulations with orbital period in the emission lines as a presence of a bright hot spot on the edge of the accretion disk.

Later on, Norton et al. [14] reported that the orbital period is (5.387±0.006) hours, corresponding to the -1 day alias of the period found by Zharikov et al. [13] and confirmed that the rotational period of the white dwarf is twice the pulse period. Soon afterwards, Zharikov et al. [15] repeated the analysis using own data along with the data provided by A. Norton. They confirmed their previous results and published the final value of $0^d.181195\pm0^d.000339$. However, they mentioned that a longer time base of observations is needed to improve this value.

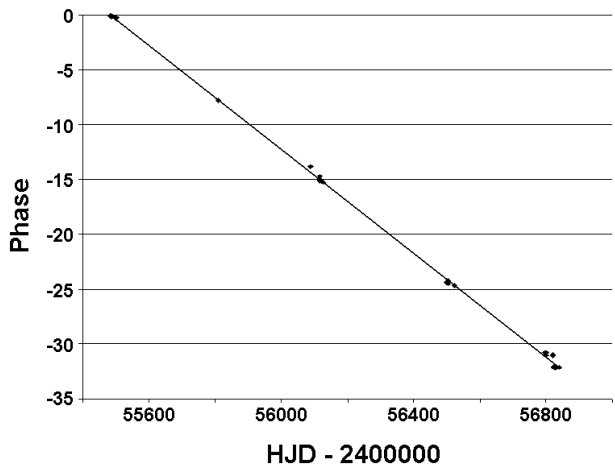 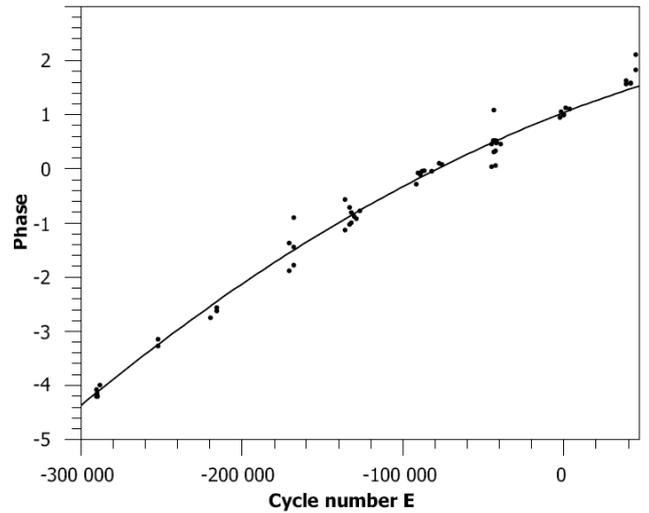

Fig. 2: (O-C) diagram of V2306 Cyg for orbital minima using data obtained in 2010-2014, the cycle miscount is well noticeable.

Fig. 3: (O-C) diagram of V2306 Cyg for spin maxima with the parabolic fit that reveals the spin-up.

In our previous work [16] we analyzed the results of 6 years of photometric monitoring of this binary system and we were not able to find spin period variations from the (O-C) analysis, but found a regular cycle miscount for the previously published orbital period and concluded that an orbital period of $0^d.181545 \pm 0^d.000003$ better fitted our data.

From the analysis of observations gathered during 9 years of photometric monitoring of the intermediate polar V2306 Cygni, we discovered its white dwarf spin period variability. The characteristic spin-up time was $(53\pm5) \cdot 10^4$ years. The value of $dP/dt = (-8.73\pm0.79) \cdot 10^{-11}$ for the spin period has an order typical of intermediate polars [11]. For the epoch of 2017, we derived the white dwarf spin rotation period of 1466.6795 seconds, with a formal accuracy of 0.0003 seconds. From the (O-C) analysis we improved the value of the orbital period of V2306 Cyg resulting in the value of $0^d.1821468 \pm 0^d.0000004$ that is 3 orders more accurate then published in [15]. It is impossible to get such high accuracy with short time base of observations.

The periodogram analysis revealed a period of 2.02 days, which was interpreted as a possible precession of the accretion disk in this system [16], but was not confirmed in our latest research.

### *RX J2133.7+5107*

The X-ray source RX J2133.7+5107 was extracted from the ROSAT Galactic Plane Survey and classified as cataclysmic variable by Motch et al. [17]. De Miguel et al. observed this object in 2010-2016. They interpreted a modulation with 6.72-h period as a negative superhump and proposed a parabolic fit to the (O-C) diagram of the spin maxima that corresponds to characteristic time of spin-up of $0.17 \cdot 10^6$ years [18].

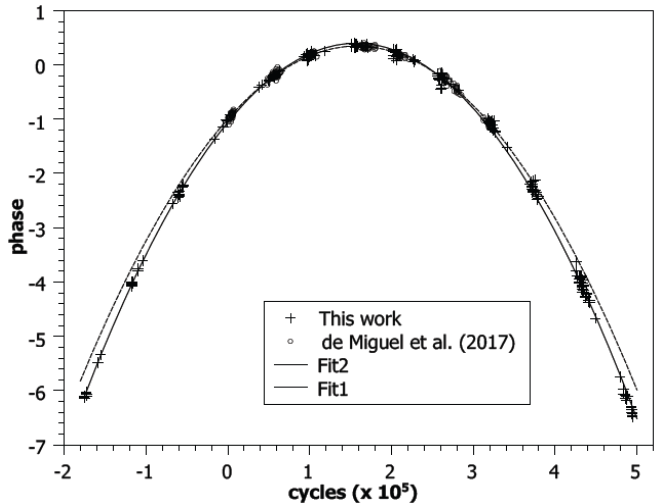

Fig. 4: (O-C) diagram of RX J2133.7+5107. The fastest spin-up rate among known intermediate polars.

From the analysis of our own observations gathered during 12 years, we determined the value of the spin-up time-scale $1.536(3) \cdot 10^5$ years. The observed rate of spin-up is even faster then reported by [18] and one of the fastest of all known intermediate polars. Statistically optimal fit to the O-C diagram was the $4^{th}$ order polynomial. This is not usual for this class of objects, so we tried to fit the O-C with a superposition of parabolic and trigonometric polynomial trend in MCV, and found a long-term variability of the spin period with a period about 7 years. The reason of these changes is a subject for discussions – either the spin rotation rate is changing near its equilibrium period, or there is a third body in this system. The negative superhumps are present in this system and the superhump period is stable from year to year.

*MU Cam*

In the case of MU Cam we have investigated the periodic modulation of the spin phases with the orbital phase (see Fig. 5). As a possible source of the unexpected scatter on this figure we have investigated the dependency of spin maxima timings on orbital phase described by Kim et al. [19].

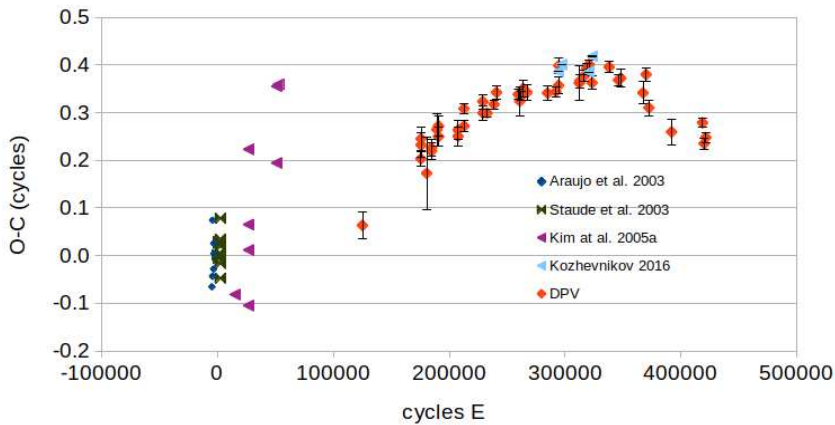

Fig. 5: (O-C) diagram of MU Camelopardalis. Notice the scatter of the plot.

As an explanation they proposed inhomogeneous accretion flow from the secondary component of the binary system. However based on our new data we can propose simple explanation as the influence of orbital sidebands in the periodic signal produced by intermediate polar. This explanation is supported by the fact that the changes in spin maxima phase are observed mainly when the sideband frequency is dominant in the periodogram.

The presence of orbital sidebands is more prominent in low states but not only. The origin of orbital sidebands can be direct accretion from the stream and/or reprocessing of X-rays at some part of the system which rotates with the orbital period.

*Other intermediate polars*

We analyzed variability of the spin period of the white dwarf in the **V405 Aur** using our observations and previously published maxima timings. As we had gaps in observational data, we presented 2 hypotheses of the spin period variability of this system - a cubic ephemeris which may be interpreted by a precession of the magnetic white dwarf or a periodic change with a period of 6.2 years and semi-amplitude of 17.2±1.8 sec [20]. The periodic variations were interpreted by a light-time effect caused by a low-mass star ($M_3=0.09M_{sun}$) [21]. Observations obtained during recent years show the continuation of descending trend of the O-C diagram. So, only first hypothesis remains active, and the cubic ephemeris should be improved using updated data set.

For the intermediate polar **EX Hya** we obtained few own time series and processed all available data from AAVSO international database and different surveys. The parabolic fit to the O-C diagram corresponds to a characteristic time scale of $4.67(12) \cdot 10^6$ years for the rotation spin-up [22], [23].

The intermediate polar **FO Aqr** shows complicated changes of the O-C diagram: according to historic data, from 1981 to 1987 the white dwarf showed a spin-down, later it changed to a spin-up [23]. The huge gap in the observational data did not allow us to fit the O-C diagram with analytical function and we fitted each part of the O-C separately [24]. Recently, we re-analyzed the maxima timings obtained between 2008 and 2018 and found the linear descending trend on the O-C, that shows us the spin rotation period during 10 years is different from that was used for O-C calculation. The residuals from the linear trend revealed sinusoidal-like changes with the period of about 16 years similar to discovered in RX J2133.7+5107 [25]. Up to now we are not able to fit the whole O-C with one smooth function, but it's ambiguous to talk about period jumps. Intermediate polars usually show smooth period variations.

*Conclusions*

Using more then 10 years of own photometric monitoring and published data we study variability of the spin periods to understand the evolution of this type of close binary systems. By this time we determined or improved the values of the spin rotation rate changes, improved values of the spin and orbital period values in selected intermediate polars and found few interesting effects in these systems. This investigation will be continued.

Magnetic cataclysmic binary systems are usually sources of polarized radiation in a wide spectral range from X-ray to radio. Despite the fact that this review includes only results obtained from optical photometry, our team studies these objects using polarimetry. S.V. Kolesnikov reviewed polarimetrical methods in astronomy [32], [33], [34], which particularly were applied for the observations of intermediate polars at the 2.6m Shain telescope of the Crimean Astrophysical observatory. For the data reduction, we have elaborated software Polarobs [35], [9].